\documentclass [12pt, a4paper]{article}

\makeatletter
\@addtoreset{equation}{section}
\makeatother
\begin{document}
\title{Nonlinear Realization of the General Covariance Group Revisited}
\author{
D. Maxera\thanks{e-mail: maxera@fzu.cz} \\
{\small \it Department of Elementary Particle Physics,}\\
{\small \it Institute of Physics of AV \v CR, Na Slovance 2, 180 40 Praha 8,
Czech Republic}}
\date{}
\maketitle
\vskip1.2cm
\noindent
\begin{abstract}
A modified algorithm for the construction of nonlinear realizations
(sigma models)
is applied to the general covariance group. Our method features
finite dimensionality of the coset manifold by letting the
vacuum stability group be infinite. No decomposition of
the symmetry group to its finite-dimensional subgroups is required.

The expected result, i.e. Einstein-Cartan gravity,
is finally obtained. 
\end{abstract}
\vspace{4.5cm}
\hfill{PRA-HEP-98-1}
 \newpage
\section{Introduction}
Nonlinear realization techniques turned out to be a useful tool
in the construction of field theories with spontaneously broken
symmetry.
As was shown in \cite{BorOgi}, also Einstein gravitation can be obtained
as a nonlinear realization of the general covariance group. In this formulation
a gravitational field is the Goldstone boson responsible for breaking
the affine group down to the Poincare group.
The method
was later successfully extended to the $N=1$ supergravity case
\cite{IN92, MN94}. 

The infinite-parametric (super)group was realized there
using the decomposition to its two finite-parametric subgroups, for which
then a standard algorithm \cite{CCWZ1, Vo73} for nonlinear realizations 
was used. However, for example Yang-Mills 
theories, where an analogical decomposition doesn't exist, were
reformulated in a nonlinear realization language without this requirement
\cite{IvaOgi76}.
Such a decomposition isn't a necessary
condition for the existence of a nonlinear realization of an infinite-parametric
group. 

Moreover, the method fails in extended supergravity cases. There it leads to
fields with too high spins without any hint how to get rid of them. This is maybe 
a more important reason for
another examination of the nonlinear realizations.

We construct a nonlinear realization
of the general covariance group by a modified method. It is based on the
straightforward realization of the whole infinite-parametric group, in
a similar way as was used in the context of $W$-algebras
\cite{Mal96} recently.
In  our approach, however, from the beginning only a finite number of Goldstone 
fields appear 
in the theory, because
most of the transformations turn out to belong to a vacuum stability group,
where they can be handled very easily.

At least two Goldstone fields enter our theory. Apart from the symmetric tensor
$h^{\mu\nu}$ there is a third rank tensor $\Gamma^{\mu\nu\rho}$ connected
to torsion. Using the inverse Higgs effect \cite{IHiggs} this 
$\Gamma^{\mu\nu\rho}$ can be further eliminated, which is related to
the vanishing of the torsion. Finally we arrive to the same invariants
as in \cite{BorOgi}.

In this way we are very close to the results obtained also in 
\cite{Pashnev}. Our approach
has the advantage of greater mathematical clarity, because we work
with finite-dimensional field manifolds entirely. Furthermore we find a
place for matter fields, as well as some other representations, in our theory.
We hope our more detailed understanding will be helpful in the
extended supergravity cases, where it is difficult to avoid too high
spins in the approach like in \cite{Pashnev} or \cite{IN92, MN94}. 

\section{Nonlinear Realization of General Covariance Group}
 
Given an infinitesimal analytic coordinate transformation 
\begin{equation}
\delta x^{\mu} = \lambda^{\mu}(x),
\end{equation}
we may regard it as a transformation encoded by Lie algebra generators
\begin{equation}
F^{n\ \nu_{1}\dots\nu_{n+1}}_{\mu}=i x^{\nu_{1}}.\dots .x^{\nu_{n+1}}
\frac{\partial}{\partial x^{\mu}}
\end{equation}
with commutation relations
\begin{eqnarray}
\left[F^{n\ \nu_{1}\dots\nu_{n+1}}_{\rho},
F^{m\ \mu_{1}\dots\mu_{m+1}}_{\kappa}\right]&=&i\sum_{i=1}^{m+1}\delta^{\mu_{i}}_{\rho}
F^{m+n\ \nu_{1}\dots\nu_{n+1}\mu_{1}\dots\mu_{i-1}\mu_{i+1}\dots\mu_{m+1}}_{\kappa}
-\nonumber\\&-&
i\sum_{i=1}^{n+1}\delta^{\nu_{i}}_{\kappa}
F^{m+n\ \nu_{1}\dots\nu_{i-1}\nu_{i+1}\dots\nu_{n+1}\mu_{1}\dots\mu_{m+1}}_{\rho}
\label{FFF}
\end{eqnarray}
Among these generators we can find
translations $P_{\mu}=F^{-1}_{\mu}$ , Lorentz transformations 
$L_{\mu \nu}=\frac{1}{2}(F^{0}_{\mu \nu}-F^{0}_{\nu \mu})$ and the special affine 
transformations  $R_{\mu \nu}=\frac{1}{2}(F^{0}_{\mu \nu}+F^{0}_{\nu \mu})$, lowering and
raising indices by flat Minkowski 
$$\eta_{\mu\nu}={\rm diag}(1,-1,-1,-1).$$

Furthermore, let us introduce the following decomposition of
the generators $F^{1}_{\mu\nu\rho}=\eta_{\mu\sigma}\eta_{\nu\kappa}F^{1\ \sigma\kappa}_{\rho}$:
\begin{eqnarray}
F^{1\ {\rm sym}}_{\nu \mu \rho}&=&\frac{1}{6}\left(F^{1}_{\nu \mu \rho}+{\rm
perm.}\right),\nonumber\\
T_{\nu \mu \rho}&=&F^{1}_{\nu \mu \rho}-F^{1\ {\rm sym}}_{\nu \mu \rho}=
\frac{1}{3}\left(2 F^{1}_{\nu \mu \rho}-F^{1}_{\nu  \rho \mu}-
F^{1}_{\rho  \mu \nu} \right).\nonumber
\end{eqnarray}

Now we are going to make a coset realization of the general covariance group G
with respect to an infinite-parametric subgroup $H$ generated
by  generators $L_{\mu \nu}$ (the Lorentz ones), 
$F^{1\ {\rm sym}}_{\nu \mu \rho}$
and all the generators $F^{n};\ n\geq 2$. This means what we are going to realize
 the group of general covariance on the coset, which can be parametrised
 by coordinates $x^{\mu}$ and tensor fields $h^{\mu \nu}(x)$
 and  $\Gamma^{\mu \nu \rho}(x)$, corresponding
to the generators $R_{\mu\nu}$ and $T_{\mu\nu\rho}$. Here  $h^{\mu \nu}$ is 
a symmetric tensor
 and $\Gamma^{\mu \nu \rho}(x)$ is a nonsymmetric tensor symmetric in the first
 two indices, i.e.
\begin{eqnarray}
\Gamma^{\mu \nu \rho}(x)&=&\Gamma^{\nu \mu \rho}(x),\label{GGG}\\
\Gamma^{\mu \nu \rho}(x)&+&\Gamma^{\nu \rho \mu}(x)+\Gamma^{\rho \nu
\mu}(x)=0.\nonumber
\end{eqnarray}
Following the algorithm of \cite{CCWZ1} 
we parametrise
the coset $G/H$ as
\begin{equation}
a(x)=e^{ixP}.e^{ih(x)R}.e^{i\Gamma(x) T}
\end{equation}
An element $g$ of the general covariance group determines a transformation
$a(x)\to a'(x')$ according to
\begin{equation}
a'(x').h'_{x}=g.a(x),\label{trC}
\end{equation}
where $h'_{x}$ is some element of the subgroup $H$. This remainder
element $h'_{x}$ can be further
decomposed as
\begin{equation}
h'_{x}=l'_{x}.q'_{x},\label{hlq}
\end{equation}
where $l'_{x}$ is an element of the Lorentz group and $q'_{x}$ is 
an element of the rest factorgroup.

There is a possibility of accommodating other fields $\psi(x)$, so
 called matter fields. They transform according to some representation
$B$ of the subgroup $H$ as
\begin{equation}
\psi'(x')=B(h'_{x})\psi(x).
\end{equation}

\section{Cartan Forms}

Now we are going to find covariant derivatives of both the Goldstone
and the matter fields. For this purpose we define Cartan forms, i.e.
 alg($G$)--valued
one-forms
\begin{equation}
a^{-1}(x)da(x)=\sum^{\infty}_{n=-1}i\omega^{n\ \mu}_{\nu_{1}\dots\nu_{n}}
F_{\mu}^{n\ \nu_{1}\dots\nu_{n}}
\end{equation}
with a nice transformation law
\begin{equation}
a'^{-1}(x')da'(x')=h'_{x}a^{-1}(x)(da(x))h'^{-1}_{x}+h'_{x}dh'^{-1}_{x}
\end{equation}

Using the commutation relations (\ref{FFF}) we can express Cartan forms in
the terms of fields $h_{\mu\nu}$, $\Gamma_{\mu\nu\rho}$.
The projections onto translations $P_{\mu}$, Lorentz $L_{\mu\nu}$ 
and special affine generators $R_{\mu\nu}$ read
\begin{eqnarray}
\omega^{\mu}_{P}&=&dx^{\nu}_{P}(r^{-1})^{\mu}_{\nu},\label{ome}\\
\omega^{\mu \nu}_{R}&=& \frac{1}{2}\{ r^{-1}(x),dr(x)\}^{\mu \nu}+
\omega_{P \rho}\Gamma^{\mu \nu\rho}\ \ {\rm and}\nonumber\\
\omega^{\mu \nu}_{L}&=&-\frac{1}{2}\left[r^{-1}(x),dr(x)\right]^{\mu \nu}-
 \omega_{P \alpha}\Gamma^{\alpha \mu \nu}
+\omega_{P \alpha}\Gamma^{\alpha \nu \mu}.\nonumber
\end{eqnarray}
with abbreviation $r^{\mu}_{\nu}(x)=\left(e^{h(x)}\right)^{\mu}_{\nu}$, i.e. a matrix
exponential of $h^{\mu}_{\nu}(x)$. The $\{ \}$, resp. $\left[\ 
\right]$
denote a matrix anticommutator, resp. commutator.

The forms corresponding
to the coset generators, i.e. $\omega_{P}$, $\omega_{R}$ and $\omega_{T}$, do not
transform 
covariantly under the transformations (\ref{trC}) by 
an element of the Lorentz group $l'_{x}$ obtained by the decomposition 
(\ref{hlq}). This is so because generally they obtain
 a supplement
from the $q'_{x}$ part of $h'_{x}$. However,
apart from  the form $\omega^{P}$ corresponding to the
translations, another exception is $\omega^{\mu \nu}_{R}$ which obtains 
an addition only
of the form $\lambda_{sym}^{\mu \nu \rho}\omega_{P \rho}$, where 
$\lambda_{sym}^{\mu \nu \rho}$ is some fully symmetric parameter.
It means the nonsymmetric part of the projection onto 
the covariant $\omega^{P}$ 
\begin{equation}
D^{\rho}h^{\mu \nu}=\frac{1}{3}\left(
2\frac{\omega_{R}^{\mu \nu}}{\omega^{P}_{\rho}}
-\frac{\omega_{R}^{\nu \rho}}{\omega^{P}_{\mu}}
-\frac{\omega_{R}^{\rho \mu}}{\omega^{P}_{\nu}}\label{cdh}
\right)
\end{equation}
does transform covariantly by a Lorentz group element.

Inserting for the Cartan forms and exploiting the symmetry properties
of $\Gamma^{\mu\nu\rho}$  (\ref{GGG}) we get
\begin{eqnarray}
D^{\rho}h^{\mu \nu}&=&\frac{1}{3}\left(
2\nabla^{\rho}h^{\mu \nu}
-\nabla^{\mu}h^{\nu \rho}
-\nabla^{\nu}h^{\rho \mu}\right)+\Gamma^{\mu\nu \rho},\label{KovGam}
\end{eqnarray}
where we have abbreviated
\begin{equation}
\nabla^{\rho}h^{\mu \nu}\equiv\frac{1}{2}r^{\rho \alpha}
\{ r^{-1},\partial_{\alpha}r\}^{\mu \nu}.\label{nabla}
\end{equation} 

\section{Covariant Derivative of the Matter Fields}
The linearly realized subgroup $H$ consists of two parts - Lorentz group
$L$ and the remaining transformations $Q$. The remainder $Q$ itself is also
a subgroup. In fact it is even a normal subgroup of $H$, though not of the whole
group of general covariance.

This normality of $Q$ inside $H$ gives us a possibility of representing it 
trivially $B(q'_{x})=1$.
Then the matter fields $\psi(x)$ are transformed only in some representation $B$ of
the Lorentz group 
\begin{equation}
\psi'(x')=B(l'_{x}).\psi(x)\label{trpsi}.
\end{equation}
A covariant derivative 
 then contains only a Lorentz algebra 
part, because the
remaining generators are effectively zero.

This means that the covariant derivative of matter fields $\psi(x)$
\begin{equation}
D^{\rho}\psi=\frac{d\psi(x)+
i\omega_{L}^{\mu\nu}B(L_{\mu\nu})\psi(x)}{\omega_{P\rho}}\label{cov}
\end{equation}
 is, exploiting the shape of $\omega_{L}^{\mu\nu}$ (\ref{ome}),
\begin{eqnarray}
D^{\rho}\psi&=&r^{\rho \alpha}\partial_{\alpha}\psi
- \left(\frac{1}{2}r^{\rho \alpha}\left[r^{-1}(x),\partial_{\alpha}r(x)\right]^{\mu \nu}
+\right.\nonumber\\
&+&\left.
\left(
\Gamma^{\rho \mu \nu}-\Gamma^{\rho \nu \mu}
\right)\right)iB(L_{\mu \nu}).\psi\label{Kovpsi}\ .
\end{eqnarray}
This covariant derivative is of course defined by its transformation properties only
up to a homogeneously transforming addition.

Equivalently, with the use of covariant derivative $D^{\rho}h^{\mu \nu}$ (\ref{cdh})
instead of $\Gamma^{\mu\nu\rho}$, we can rewrite
\begin{eqnarray}
D^{\rho}\psi&=&r^{\rho \alpha}\partial_{\alpha}\psi
- \left(\frac{1}{2}r^{\rho \alpha}\left[r^{-1}(x),\partial_{\alpha}r(x)\right]^{\mu \nu}
-\frac{1}{2}r^{\nu \alpha}\{r^{-1}(x),\partial_{\alpha}r(x)\}^{\mu \rho}+
\right.\nonumber\\
&+&\left.\frac{1}{2}r^{\mu \alpha}\{r^{-1}(x),\partial_{\alpha}r(x)\}^{\nu \rho}+
D^{\nu}h^{\rho \mu}-D^{\mu}h^{\rho \nu}\right)iB(L_{\mu \nu}).\psi\ .
\end{eqnarray}

We leave here open the question of that representations of the
vacuum stability group $H$, where the $Q$-part doesn't act trivially.
However, as we will show in the next section, known matter fields
fit to the above considered case.

\section{Interpretation in Terms of Differential Geometry}

We have obtained a theory invariant with respect to general coordinate
transformations. The field content is apart from matter fields
a symmetric tensor $h_{\mu \nu}$ and a tensor of the third rank
$\Gamma_{\mu \nu\rho}$ with the symmetry properties (\ref{GGG}).
This reminds us slightly of the formulation of gravity
in terms of a metric $g_{ab}$ 
and an affine connection $\kappa^{a}_{bc}$.

The relation of our formulation to the classical one can be established
as follows. First of all let us mention that the group of stability
of the origin $G_{o}$, which includes all the transformations of the general covariance
group up to translations, has a covariant vector representation $R_{v}$. By this we mean
the representation defined as the ordinary matrix representation for $GL(4)$ part
of $G_{o}$, i.e.
\begin{eqnarray}
\left(R_{v}\left(L^{\mu}_{\ \nu}\right)\right)^{\beta}_{\ \alpha}&=&-\frac{i}{2}
\left(\delta^{\mu}_{\alpha}\delta^{\beta}_{\nu}-
\eta^{\mu\beta}\eta_{\nu\alpha}\right),\label{mar}\\
\left(R_{v}\left(R^{\mu}_{\nu}\right)\right)^{\beta}_{\alpha}&=&-\frac{i}{2}
\left(\delta^{\mu}_{\alpha}\delta^{\beta}_{\nu}+
\eta^{\mu\beta}\eta_{\nu\alpha}\right),\nonumber
\end{eqnarray} 
and all the other generators represent trivially
\begin{equation}
F^{n\ \nu_{1}\dots\nu_{n+1}}_{\mu}=0\ \ {\rm for}\ n\geq 1.
\end{equation}
In an analogical way other tensor representations can be defined.

Now let us return to the transformation law (\ref{trC}). We can rewrite
$g.a(x)\equiv g.e^{ixP}.e^{ih(x)R}.e^{i\Gamma(x) T}$ as
\begin{eqnarray}
g.e^{ixP}.e^{ih(x)R}.e^{i\Gamma(x) T}=e^{ix'P}.g_{o}(x, g).e^{ih(x)R}.e^{i\Gamma(x) T},
\end{eqnarray}
where $g_{o}(x, g)$ is just an element of the stability group of origin $G_{o}$ 
determined by $g$ and $x$.

This element $g_{o}(x)$, taken in the covariant vector representation, is exactly 
$\frac{\partial x'}{\partial x}$
\begin{equation}
\left(R_{v}\left(g_{o}(x)\right)\right)^{\mu}_{\nu}=
\frac{\partial x'^{\mu}}{\partial x^{\nu}}.\label{Rvgo}
\end{equation}
This is a key relation in the relationship between the two formulations.

The proof is very straightforward. First we are going to calculate
 $g_{o}$ for
an infinitesimal transformation and
then insert the vector representation. The calculation of 
$g_{o}$ can best be done in the original representation (\ref{FFF}),
using however a variable $y$ instead of $x$ to prevent mixing it up with
the coset coordinate. Then we obtain
\begin{eqnarray}
&&g.e^{ixP}=e^{-\lambda^{\mu}(y)\frac{\partial}{\partial y^{\mu}}}.
e^{-x^{\nu}\frac{\partial}{\partial y^{\nu}}}=
e^{-x^{\nu}\frac{\partial}{\partial y^{\nu}}}.
e^{-\lambda^{\mu}(y+x)\frac{\partial}{\partial y^{\mu}}}=\\
&=&
e^{-(x^{\nu}+\lambda^{\mu}(x))\frac{\partial}{\partial y^{\nu}}}.
e^{-\left(\left.
\frac{\partial \lambda^{\mu}(y+x)}{\partial y^{\rho}}\right|_{y=0} y^{\rho}
\frac{\partial}{\partial y^{\mu}}\right)
}+o(\lambda^{2})+o(y^{2})=\nonumber\\
&=&
e^{-(x^{\nu}+\lambda^{\mu}(x))\frac{\partial}{\partial y^{\nu}}}.
e^{-
\frac{\partial \lambda^{\mu}(x)}{\partial x^{\rho}} y^{\rho}
\frac{\partial}{\partial y^{\mu}}
}+o(\lambda^{2})+o(y^{2})\nonumber
\end{eqnarray}
From this it follows that in the covariant vector representation (\ref{mar})
\begin{equation}
\left(R_{v}\left(g_{o}(x)\right)\right)^{\mu}_{\nu}=\delta^{\mu}_{\nu}
+\frac{\partial \lambda^{\mu}(x)}{\partial x^{\nu}},
\end{equation}
which is exactly the key relation (\ref{Rvgo}).

Now we are able to identify tetrades and other quantities according
to their transformation properties.
We observe that coset elements  in the covariant
vector representation $r^{\mu}_{\nu}(x)=R_{v\ \nu}^{\mu}(e^{ih(x)R})$
transform as
\begin{equation}
r'^{\mu}_{\nu}(x')=
\frac{\partial x'^{\mu}}{\partial x^{\rho}}
r^{\rho}_{\lambda}(x)
\left(R_{v}(l'^{-1}_{x})\right)_{\nu}^{\lambda}.
\end{equation} 
This means that although $r^{\mu \nu}$ is a symmetrical matrix,
each of its indices transforms differently. Thus we arrive at two
types of indices - holonomic, resp. anholonomic ones, i.e. those transforming with 
$\frac{\partial x'^{a}}{\partial x^{b}}$ and  those transforming
by the Lorentz transformation $\left(R_{v}(l'^{-1}_{x})\right)_{\nu}^{\lambda}$,
respectively.
From now on we will carefully distinguish these two types by denoting
the holonomic indices by latin letters and
the anholonomic ones by greek letters. Thus we have
$r^{a}_{\mu}$ and $r^{-1\mu}_{a}$

If there is a matter field $\psi^{\nu}(x)$ transforming
as a covariant 
vector with respect to the Lorentz group elements $l'_{x}$,
the quantity $\psi^{a}= r^{a}_{\nu}(x)\psi^{\nu}(x)$ is a covariant
holonomic vector.
Similarly, we can construct a holonomic contravariant vector
as $\psi_{a}=r^{-1\lambda}_{a}\psi_{\lambda}$. This suggest that
$r^{a}_{\mu}$ plays the role of a tetrade.

Using the fact that any tensorial (i.e., non--spinorial) representation of the
Lorentz group can be extended to the representation of
the whole $GL(4)$, we can construct a holonomic tensor
$t^{a_{1}\dots a_{m}}_{b_{1}\dots b_{n}}$ from a
given matter field $t^{\mu_{1}\dots\mu_{m}}_{\nu_{1}\dots\nu_{n}}$
and the
coset element $e^{ih(x)R}$ in a proper representation analogously. 
One arrives at
\begin{equation}
t^{a_{1}\dots a_{m}}_{b_{1}\dots b_{n}} =
r^{a_{1}}_{\mu_{1}}\dots r^{a_{m}}_{\mu_{m}}
r^{-1\nu_{1}}_{b_{1}}\dots r^{-1 \nu_{n}}_{b_{n}}
t^{\mu_{1}\dots\mu_{m}}_{\nu_{1}\dots\nu_{n}}.\label{Trrt}
\end{equation}

Furthermore, the quantities $g_{ab}=r^{-1\mu}_{a}r^{-1 \nu}_{b}\eta_{\mu\nu}$,
resp. their inverse 
$g^{ab}=r_{\mu}^{a}r_{\nu}^{b}\eta^{\mu\nu}=
(g_{ab})^{-1}$ determine an invariant scalar product of covariant,
resp. contravariant vectors and thus
play the role of a metric field. 

We can conclude that
quantity $r$, the coset element in a vector representation,
plays the role of a tetrade which is preserved symmetric by a
proper the Lorentz transformation of the anholonomic index when subject to
a general covariance transformation.

It remains to answer what are the counterparts
of the covariant derivatives (\ref{KovGam}), (\ref{Kovpsi})
and, related to this, what is the
role of the field $\Gamma^{\mu \nu \rho}$

The covariant derivative $D^{\rho}\psi^{\mu}$ (\ref{Kovpsi}) transforms
as a second rank anholonomic tensor. The corresponding holonomic tensor 
$\Delta^{a}\psi^{b}=r^{a}_{\rho}r^{b}_{\mu}
D^{\rho}\psi^{\mu}$ can be rewritten, using the formula (\ref{Kovpsi})
for the covariant derivative  and the abbreviation (\ref{nabla}), as
\begin{equation}
\Delta_{a}\psi^{b}=r^{-1}_{a\rho}r^{b}_{\mu}
D^{\rho}\psi^{\mu}=\partial_{a}\psi^{b}+\kappa_{ac}^{b}
\psi^{c}
\end{equation}
where, with help of the identity 
$\nabla^{\rho}h^{\mu\nu}=-\partial_{a}g_{bc}r^{a\rho}r^{b\mu}r^{c\nu}$,
\begin{eqnarray}
\kappa_{ac}^{b}
&=&-\left(\nabla^{\rho}h^{\mu \nu}
-\Gamma^{\rho \mu \nu}+\Gamma^{\rho \nu \mu}\right)
r^{-1}_{c\nu} r^{b}_{\mu}r^{-1}_{a\rho}=\\
&=&\frac{1}{2}g^{bd}(-\partial_{d}g_{ac}+\partial_{a}g_{cd}+\partial_{c}g_{ad})+
\left(D^{\nu}h^{\rho \mu}-D^{\mu}h^{\rho \nu}\right)r^{-1}_{c\nu} r^{b}_{\mu}r^{-1}_{a\rho}.\nonumber
\end{eqnarray}

Similarly, for each tensorial
representation 
\begin{eqnarray}
\Delta_{c}t^{a_{1}\dots a_{m}}_{b_{1}\dots b_{n}} &\equiv&
r^{a_{1}}_{\mu_{1}}\dots r^{a_{m}}_{\mu_{m}}
r^{-1\nu_{1}}_{b_{1}}\dots r^{-1 \nu_{n}}_{b_{n}}
r^{-1}_{c\rho}D^{\rho}t^{\mu_{1}\dots\mu_{m}}_{\nu_{1}\dots\nu_{n}}=\\
&=&\partial_{c}t^{a_{1}\dots a_{m}}_{b_{1}\dots b_{n}}+
\sum^{n}_{i=1}\kappa_{cd}^{\ a_{i}}
t^{a_{1}\dots a_{i-1}da_{i+1}\dots a_{m}}_{b_{1}
\dots b_{n}}
-\sum^{m}_{j=1}\kappa_{cb_{j}}^{\ d}
t^{a_{1}\dots a_{m}}_{b_{1}
\dots b_{j-1}db_{j+1}
\dots b_{n}}.\nonumber
\end{eqnarray}

Let us note that it implies together with 
$D^{\rho}\eta^{\mu\nu}=D^{\rho}\eta_{\mu\nu}=0$
that
\begin{equation}
\Delta_{a}g^{bc}=\Delta_{a}g_{bc}=0.
\end{equation}
This means that nonmetricity is essentially zero in this formulation.

Furthermore we observe that the connection $\kappa_{ac}^{b}$
isn't symmetric. The torsion $S_{ac}^{b}$, 
i.e. the part of $\kappa_{ac}^{b}$
antisymmetric in $a$, $c$, is then
\begin{eqnarray}
S_{ac}^{b}=\kappa_{ac}^{b}-\kappa_{ca}^{b}
=\left(D^{\nu}h^{\rho \mu}-D^{\rho}h^{\nu \mu}\right)
(r^{-1})_{c\nu} r^{b}_{\mu}r^{-1}_{a\rho}.\label{tor}
\end{eqnarray}

Let us mention that there is a possibility of another choice of the subgroup
$H$. We could leave some more generators in the "broken" (coset) part of the
general covariance group $G$. For example leaving there the generator
$F^{1\ sym}_{\mu\nu\rho}$, $\kappa^{b}_{ac}$ would also yield the fully symmetric
part, thus obtaining all the degrees of freedom of the
connection. However, considering the transformation properties only,
this additional nonlinearly transforming field would be superfluous,
because it can be expressed in terms of the metric and a linearly transforming
field.

\section{Elimination of $\Gamma^{\mu \nu \rho}$}

The covariant derivative of the field $h^{\mu\nu}$ (\ref{KovGam})
 gives us the possibility of expressing $\Gamma^{\mu \nu \rho}$
in terms of $h_{\mu \nu}$ and its derivatives via so called inverse Higgs
effect, i.e. fixing $D^{\rho}h^{\mu \nu}$ to be zero
\begin{equation}
D^{\rho}h^{\mu \nu}=0.\label{Dh0}
\end{equation}
 Then we have
\begin{equation}
\Gamma^{\mu\nu \rho}=-\frac{1}{3}\left(
2\nabla^{\rho}h^{\mu \nu}
-\nabla^{\mu}h^{\nu \rho}
-\nabla^{\nu}h^{\rho \mu}\right).\label{Gam}
\end{equation}

This elimination can be inserted into the Cartan form $\omega^{\mu \nu}_{L}$ 
corresponding to the Lorentz
generators $L_{\mu \nu}$, which affects
the covariant derivative (\ref{cov}) of matter fields $\psi(x)$.
\begin{eqnarray}
D^{\rho}\psi&=&r^{a \rho}\partial_{a}\psi
- \left(r^{a \rho}\left[r^{-1}(x),\partial_{a}r(x)\right]^{\mu \nu}
+\right.\nonumber\\
&+&\left.
\left(\nabla^{\mu}h^{\rho \nu}-
\nabla^{\nu}h^{\rho \mu}\right)\right)iB(L_{\mu \nu}).\psi\label{elcov}\ .
\end{eqnarray}

The condition (\ref{Dh0}) is equivalent to vanishing of the torsion. 
Really, from the shape of the torsion 
(\ref{tor}) and the symmetries of $D^{\rho}h^{\mu \nu}$
it follows that
\begin{equation}
D^{\rho}h^{\mu \nu}=(r^{-1})^{\mu}_{b}r^{a \nu} r^{c \rho}S^{b}_{ac}
+(r^{-1})^{\nu}_{b}r^{a \mu} r^{c \rho}S^{b}_{ac}.
\end{equation}
which together with (\ref{tor}) gives
\begin{equation}
D^{\rho}h^{\mu \nu}=0 \Longleftrightarrow S^{a}_{bc}=0
\end{equation}

\section{Invariant Action}

In order to construct a suitable action we need a list of invariants of the
theory. First of all we construct an invariant volume, which can be
made of the Cartan forms $\omega_{P}$ corresponding to the translations
\begin{equation}
\omega_{V}=\omega^{0}_{P}\wedge\omega^{1}_{P}\wedge\omega^{2}_{P}\wedge
\omega^{3}_{P}={\rm det}( r^{a}_{\mu})^{-1}d^{4}x.
\end{equation}

The invariant Lagrangian for the matter fields can be constructed by replacing
ordinary derivatives by the covariant ones (\ref{elcov}) in their Lorentz
invariant Lagrangian. It remains to find a Lagrangian for the gravity
field itself.

There is a curvature $R^{a}_{bcd}$ which can be defined as a commutator of 
the covariant 
derivatives
of a vector.
This completes the list of invariants in terms of the tetrade field 
$r^{a}_{\mu}$ without more than two
derivatives. 

One may wonder if it would not be possible to construct a well-transforming
object by adding to the Cartan form $\omega_{T}^{\mu\nu\rho}$ some suitable
combination of the other forms. This is indeed possible, the
result however is again the curvature because of the Cartan equations.

Anyway, we have all the needed 
components of the gravitational Lagrangian.

\section{Conclusions}

We have rederived Einstein--Cartan theory using a modified non-linear realization
algorithm, in which from the beginning only a finite number of generators of the
general covariance group are broken. The symmetric tensor field arising as a
Goldstone boson responsible for breaking the special affine generators
is finally the only essential nonlinearly-transforming field in
the theory. 

The method used doesn't require an infinitely--parametric symmetry to be
decomposable onto  finite-parametric subgroups, so it is more general
then the previously used algorithm.
\vspace{0.5cm}

{\large \bf Acknowledgements}\\
\\
I wish to thank J. Niederle and M. Bedn\'{a}\v{r} for useful discussions.


\end{document}